# Locating earthquakes with a network of seismic stations via a deep learning method

Xiong Zhang[1], Jie Zhang[1]*, Congcong Yuan[1], Sen Liu[2], Zhibo Chen[2], Weiping Li[2]

[1]*School of Earth and Space Sciences, University of Science and Technology of China, Hefei, Anhui 230026, P. R. China.*

[2]*School of Information Science and Technology, University of Science and Technology of China, Hefei, Anhui 230026, P. R. China.*

**Abstract:** The accurate and automated determination of earthquake locations is still a challenging endeavor. However, such information is critical for monitoring seismic activity and assessing potential hazards in real time. Recently, a convolutional neural network was applied to detect earthquakes from single-station waveforms and approximately map events across several large surface areas. In this study, we locate 194 earthquakes induced during oil and gas operations in Oklahoma, USA, within an error range of approximately 4.9 km on average to the epicenter and 1.0 km to the depth in catalogs with data from 30 network stations by applying the fully convolutional network. The network is trained by 1,013 historic events, and the output is a 3D volume of the event location probability in the Earth. The trained system requires approximately one hundredth of a second to locate an event without the need for any velocity model or human interference.

## 1. INTRODUCTION

Locating earthquakes constitutes a fundamental problem in seismology (*1-3*). In particular, the reporting of earthquake locations (or hypocenters) in real time helps provide an assessment of potential hazards in local areas. For moderate to large earthquakes, such real-time reporting could lead to the issuance of early warnings to the public prior to the arrival of destructive and deadly seismic waves; for small earthquakes, it helps characterize subsurface activities and delineate fault movements. An earthquake occurs when two blocks within the earth suddenly slip past one another. In addition to tectonism, seismicity can be induced by the addition or removal of either surface water or groundwater and by the injection or removal of fluids due to industrial activity (*4-8*). For example, approximately 900 widely felt M ≥ 3 earthquakes occurred in north-central Oklahoma in 2015, while only one M ≥ 3 earthquake occurred in Oklahoma on average each year before 2009 (*8*). It is now widely recognized that this almost 900-fold increase in earthquake occurrence is related to the widespread disposal of saltwater being coproduced with oil in seismically active areas (*8*). Therefore, there is a strong demand for technology that can timely and accurately report earthquakes automatically, as such information may immediately affect industrial activities and the actions of local residents near earthquake epicenters.

Earthquakes are conventionally located through a process composed of detecting events, picking the arrival times of P-waves, and estimating the hypocentral parameters from the arrival times using a velocity model. Picking the first arrivals may also serve as event detection. Moreover, picks of P-wave arrival times from two or more seismic stations are needed to locate an event. Utilizing arrival times to locate earthquakes as opposed to waveforms simplifies the problem considerably; the corresponding methods, which include travel time inversion (*9*), grid search (*10, 11*), and double-difference techniques (*12*), are implemented in many different forms (*13-16*). However, conventional arrival time methods suffer from uncertainties in the time picks, inaccurate velocity models, and non-unique solutions. Thus, human interference and/or confirmation are often needed to avoid false results.

Three-component earthquake waveform data should contain more earthquake information than only the arrival times of P-waves. However, most waveform studies performed to date have focused on event detection problems (*17-22*), although the ultimate goal of earthquake reporting is to determine the hypocenter, magnitude, and origin time. Furthermore, utilizing waveform data to locate earthquakes in real time is challenging because numerous parameters influence seismic data in addition to the hypocentral parameters, and the numerical computations may be costly as well. Among the few efforts to develop an automated detection system, an earthquake search engine method that applies fast search algorithms in computer science was introduced to find the best match for an earthquake waveform from a preset synthetic database, thereby returning the source information from the matched synthetic within a second (*23*). This method is robust for dealing with long-period data at a large recording scale, but it is difficult to implement for regional or local earthquake monitoring, since the waveform data for which are highly sensitive to structural heterogeneities. Recently, another attempt was performed to apply artificial intelligence, specifically, the convolutional neural network (CNN) method, to detect seismic events from streaming waveform data (*21*). This method can detect more than 17 times more earthquakes than a catalog by using single-station data in real-time applications, and it also outputs the probabilistic locations of detected events. However, CNN methods that implement the multilabel classification of training data from single-station waveforms could only approximately map induced seismicity in Oklahoma across six large areas. Unfortunately, while these probabilistic surface locations are helpful, they are not comparable to the hypocenter accuracy required for earthquake catalogs (*24*).

In this study, we focus on real-time earthquake location problems by accessing seismic waveform data from a regional network of 30 stations in Oklahoma. This study assumes that earthquake events have already been detected from the network data, and those events are selected to further determine the event epicenter and depth. Motivated by the recent success of applying CNNs to solve inverse problems in medical imaging (*25*), we design a novel architecture, namely, the fully convolutional network (FCN), which can predict a 3D image of the earthquake location probability in the Earth from a volume of raw input data recorded at multiple network stations. This approach is different from the typical application of CNNs to classification tasks, where the output for the input data is a single class label. Instead, the output of our network includes a large number of pixels representing a 3D image, in which the peak value corresponds to the most likely source location in the Earth. Similar efforts for representing ground truth with image pixels have been made in image segmentation (*26, 27*), medical image reconstruction (*28*), and synthesizing high-quality images from text descriptions (*29*). A deep learning approach for earthquake location is appealing because it overcomes many limitations of inversion methods; there is no need to handcraft parameters for forward modeling, objective function, regularization, and optimization. The FCN performance is robust even in the case of a limited amount of training data. To monitor the induced seismicity in Oklahoma, 1,013 historic earthquakes are used as training samples, and 194 events are used for testing. The method requires approximately one hundredth of a second to locate an event, which is generally within an average range of 4.9 km to the epicenter and 1.0 km to the depth in an earthquake catalog.

## 2. RESULTS

### 2.1. Data

The sharp increase in the occurrence frequency of small- to moderate-sized earthquakes in Oklahoma, USA, since 2009 has drawn elevated concerns regarding the potential for earthquake hazards in this area (*4, 5, 30, 31*). Many studies have shown that the sharp increase in seismicity in Oklahoma is principally caused by the large-scale injection of saltwater into the Arbuckle group (*7, 30, 32, 33*). To monitor the induced seismicity in Oklahoma, the temporary Nanometrics Research Network consisting of 30 broadband seismic stations operated by Nanometrics Seismological Instruments was deployed in this region from 10 June 2013 to 31 March 2016. The minimum station interval varies from 14 to 30 km, and the signals are recorded from 0.1 Hz to 30 Hz on all three components. We selected 1,207 events with seismic moment magnitudes ranging from Mw 3.0 to Mw 4.9 as cataloged by the U.S. Geological Survey and divided the events into two groups: one group of 1,013 events is used to train the neural network, and the other 194 events are utilized to test the trained model. To simulate the real situation, we use the early events as training samples (from 10 June 2013 to 6 November 2015) and the latest events as testing samples (from 7 November 2015 to 31 March 2016). In a later section, different numbers of events in each group are also tested to study the performance of the neural network. The seismic network covers an area of approximately 320 km × 270 km in Oklahoma. In this study, we assume that the earthquakes have already been detected and that the corresponding waveforms are truncated in a time window from the continuous records. Without any processing applied to the data, we obtain the hypocenter solutions by employing the FCN model with the raw input waveforms from 30 stations.

### 2.2. 3D location image

In machine learning problems, we need to pair the input data and the output results in a quantitative manner. Due to the underlying physics, there is a nonlinear relationship between the seismogram data and event location parameters in an earthquake location problem. Accordingly, instead of generating a single class label for the earthquake location, our FCN model outputs a 3D image volume that represents the probability of the event location in the subsurface, as shown in the bottom plot of Fig. 1A. The point within the image with the largest magnitude marks the most likely event location. The details of the network architecture illustrated in Fig. 1A will be elaborated in the section of Methods. Through numerical studies with a grid search method to calculate the misfit of the arrival times of an event in the subsurface, we find that the distribution of the misfit somewhat reflects the probability of the event location, where the minimum misfit corresponds to the most likely location. A Gaussian function with an appropriate radius is sufficient to approximately represent the probability, and the radius parameter of the Gaussian function affects the peak location and the value of each pixel in the output. If the radius is excessively large or small, the testing errors will increase; therefore, an optimal value should be determined through a few tests prior to training the network. The design of the Gaussian function for representing event location is critical for reducing the number of training samples required. This is because the training set pairs with all of the pixels in the Gaussian function collectively instead of each pixel independently. The pixels of a Gaussian function are constrained with one another in the training process.

Our ground truth of the event location for the training is represented by a 3D Gaussian function, the peak point of which represents the event location, and the peak value is 1.0. However, testing examples with new data reveal that the output may not maintain the shape of a Gaussian function, and the peak value may vary depending on the uncertainty in the result. As shown in one of the examples below, the peak value will be significantly lower if the true event location of the testing data is outside the 3D volume. Moreover, the result is expected to be more accurate for higher peak values. Therefore, we are able to eliminate false results using a preset threshold of the probability value. False results may include events outside the interest zone or waveform data that do not significantly resemble any event in the training dataset.

To monitor induced seismicity in Oklahoma, the volume range of our output is constrained by our zone of interest bounded by the latitude range from 34.975° to 37.493°, the longitude range from -98.405° to -95.527°, and the depth range from 0 km to 12 km. As designed in our network, the output volume is 80 × 128 × 30 grids, representing a study area with dimensions of 2.518° × 2.878° × 12 km. This means that the grid spacing in the 3D pixel volume is 0.0315° in latitude, 0.0225° in longitude, and 0.4 km in depth. As described above, we establish a Gaussian distribution in the study zone with the peak point situated at the true location. The ground truth is obtained from the U.S. Geological Survey earthquake catalog. The training is performed on a graphics processing unit (GPU; GeForce GTX 1070) with three-component data for 1,013 events from 30 stations, and the training process is completed in approximately 3 hours. We utilize 200 epochs with a batch size of 4 to train the network. Fig. 1B and 1C show the event epicenters of the 1,013 training samples and the convergence of the loss function, which approaches zero after 10 epochs. Consequently, the trained model in the final epoch is ready to predict the locations for the new data.

### 2.3. Testing with new data

To assess the location performance of our deep learning algorithm, we test the 194 latest events in Oklahoma with the FCN model. The network takes in three-component data recorded at 30 stations in the form of the RGB color model and produces an output consisting of a 3D location image for each event. Fig. 2 shows a testing example with an event that occurred on 31 March 2016, with a magnitude of Mw 3.2. The input data are displayed in Fig. 2A, and the predicted 3D location image is shown in Fig. 2B. The value of the peak point is 0.9, and the predicted location is approximately 2.5 km away from the ground truth marked by the white star (Fig. 2C). In this example, the data of station 29 show enormously large noise and the instrumentation of station 6 might incorrectly function as well. However, the results are not nearly affected. This is one of the advantages using data from a network as opposed to a single station. It also demonstrates the benefits of applying a deep learning method rather than precise calculation in conventional geophysics to deal with noisy data.

Fig. 3 presents the location results for all of the 194 testing events; the ground truth is illustrated in Fig. 3A, and the testing results are provided in Fig. 3B. Fig. 3C shows the relative epicenter errors to the ground truth with an average of 4.9 km, and Fig. 3D shows the relative depth errors to the ground truth with an average of 1.0 km. The ground truth of these 194 testing events is also obtained from the earthquake catalog produced by conventional manual processing; conventional event catalogs may include calculation errors due to various factors, including a simplified velocity model and methodology as well as uncertainties in the time picks. Therefore, our testing errors are relative but within the error range of typical manual location results at such a regional scale (320 km × 270 km) (*34, 35*). Furthermore, without any human interference, our approach is fully automated and fast, that is, each location problem requires approximately one hundredth of a second to complete.

We further evaluate the performance of the FCN model with a number of different training samples. We use the same 194 testing events to calculate the mean errors of the predicted locations for the FCN model trained with a different number of samples. As shown in Fig. 4, the epicenter errors decrease with an increase in the number of training samples. The depth errors are generally small over the different number of training samples; this may be because most of the training events are in the depth range from 4 to 7 km in Oklahoma. With approximately 1,000 training events, the location errors seem acceptable and the error curve on Fig. 4a suggests more training samples may continue to improve the results. A training set with about 1,000 samples is considered a very small amount of training data in deep learning applications. If we apply the CNN classification method to solve the earthquake location problem with a similar resolution and take each possible location pixel as a class, several hundreds of thousands of classes are needed, and thus, an enormous number of samples would be required to train the network for such a large-scale classification problem. In an image classification example with 1,000 classes, approximately 1.2 million images are required for training (*36*).

### 2.4. Preventing false results

Our deep learning network is designed to monitor induced seismicity in Oklahoma. What will it happen if an event originates from outside the interest zone? To address this concern, we design a test with an earthquake occurring outside of Oklahoma, as shown in Fig. 5. This Mw 4.0 event (blue star) occurred at 37.429°, -98.954° on 23 May 2015, and it was approximately 123 km away from the nearest seismic station of the network. Using our FCN model, the event is located within the upper-left corner of Oklahoma. This incorrect location is anticipated because all of the output from our FCN model should consist of events that are located inside the preset 3D volume. However, in this case, the peak value of the output 3D image is only 0.5, which is much smaller than the peak values for all of the events occurring inside the 3D volume (>0.9), suggesting a very low location probability. Therefore, we can eliminate and prevent false results on the basis of the peak value using a preset threshold.

## 2.5. Potential for early warning

The application of deep learning to the real-time location of seismic events could enable the implementation of an earthquake early warning system. Earthquake early warning systems are intended to issue warnings to the public between a few seconds and slightly over 1 min after an event occurs (*37*). Such efforts require locating seismic events with only partial incoming data as quickly as possible and updating the results as additional data for the same event become available.

Several strategies should be tested and evaluated with partial input data using our deep learning method. One simple test is to assess the performance of the proposed FCN model trained with complete data samples but tested with partial input data. The size of the input dataset is fixed in the network, but one can keep the partial initial data and set the rest to zero in the input. However, this test shows that the resulting accuracy is very low because the testing samples are unfamiliar with the trained network. Another test is to train the network with partial data samples within four windows at fixed times, for example, 0-5 s, 0-30 s, 0-55 s and 0-90 s, and to keep the testing data in the same window sizes. Note that the zero time of these windows is the time for the very first arrival of the event recorded by a station. We also train the network with partial data samples within random window lengths and then test the partial data with arbitrary time lengths. The random window length for the four windows varies over different training samples. In the Supplementary Information, we provide detailed documentation on testing procedure and results. In summary, the last strategy with random window lengths for the training samples produces the highest accuracy in locating earthquakes with partial initial data. This strategy augments the size of our training set by a factor of 4, and the use of random window lengths enhances the prediction capability for input data with an arbitrary length. Fig. 6 shows the prediction results for an event with three different time windows. The initial arrival of the seismic event was recorded at 5, 12, and 29 stations, as shown in Fig. 6A, 6B and 6C, respectively, and the estimated location of the epicenter is improved with the availability of more data.

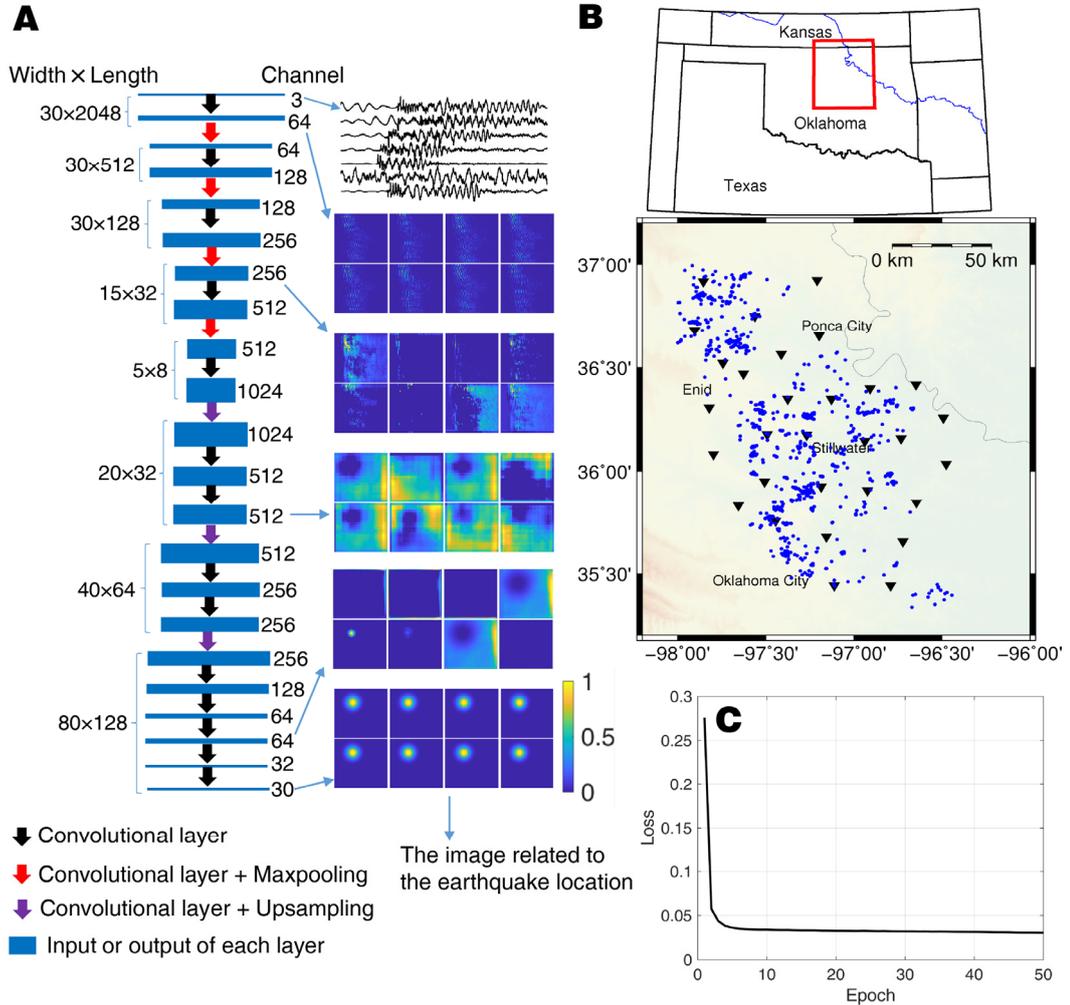

**Fig. 1. The neural network architecture and network training.** (**A**) The sizes of the input and output data are labeled on the left side of the network architecture, and the depth (channel) of the data is labeled to the right of the network architecture. We list only the images of the eight selected channels for each output from some of the layers. (**B**) The red box for the region of interest, and 1,013 historical events selected in the training set. (**C**) The loss curve during the training.

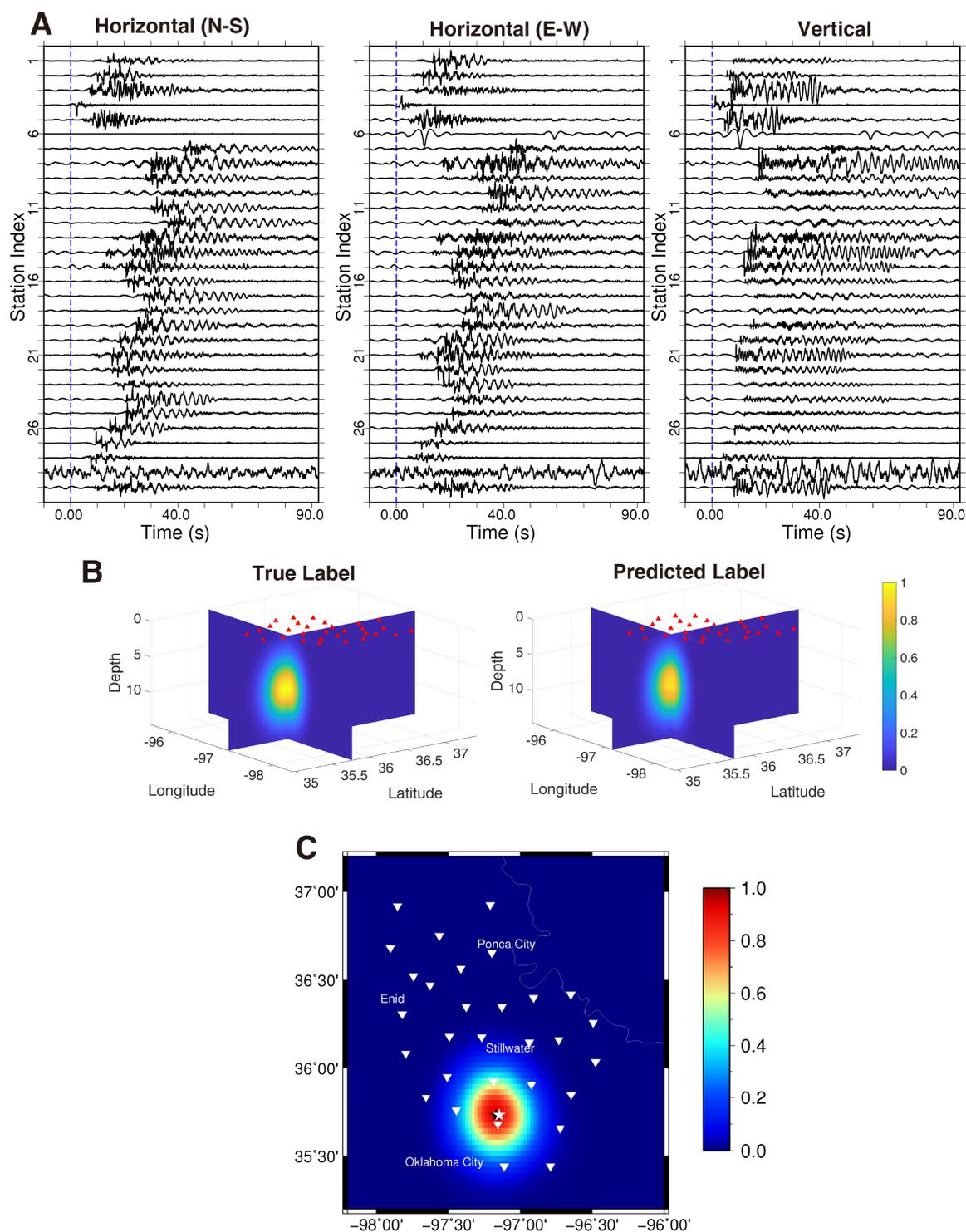

Fig. 2. **The prediction results for a testing event.** (**A**) The three components of the input waveform for an earthquake in 2016. (**B**) The true and predicted labels; the red triangles denote seismic stations. (**C**) The true and predicted locations; the white triangles denote seismic stations; the black and white stars are the predicted and true locations, respectively; the color image with magnitude for probability in panel C shows a 2D section extracted from the 3D volume in panel B.

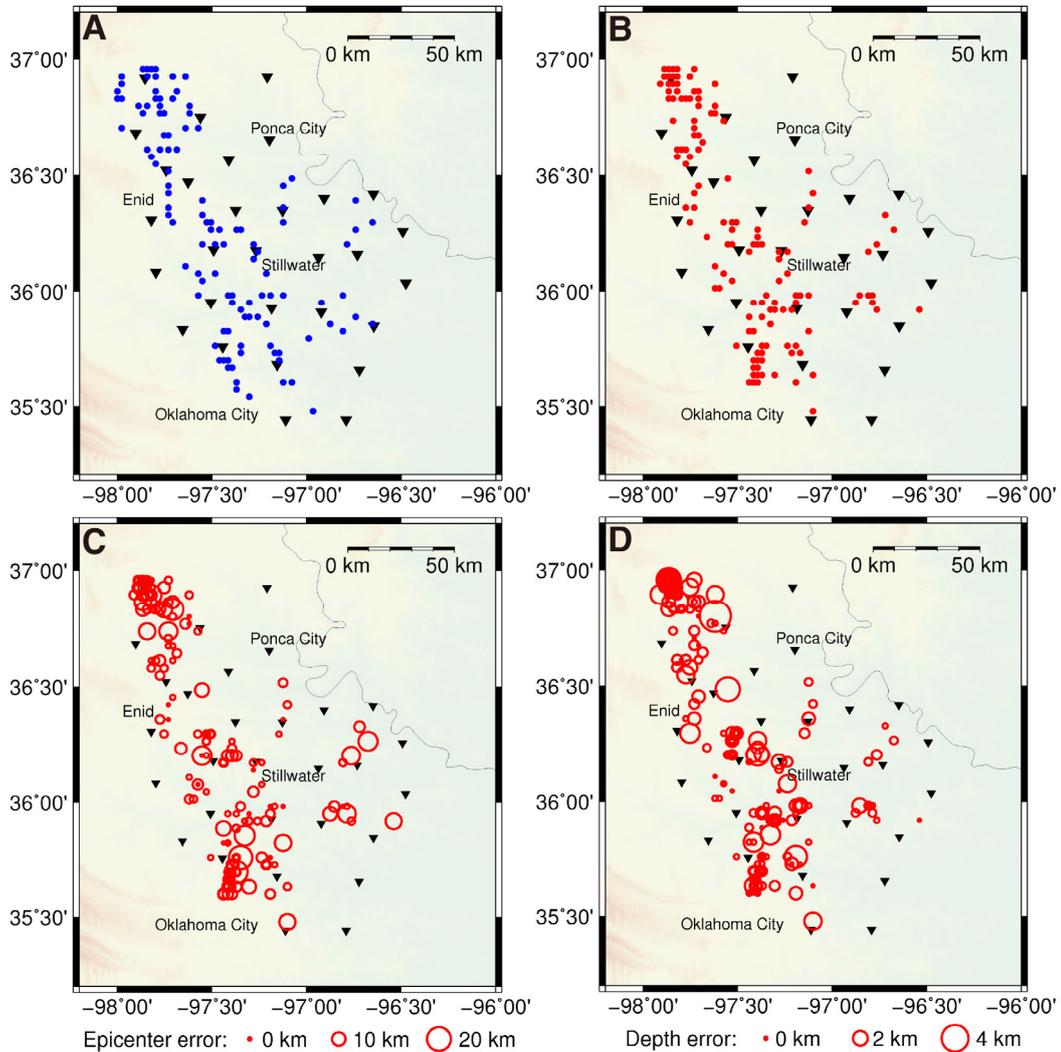

**Fig. 3. The prediction results for 194 testing events.** (**A**) The true epicenters of the 194 testing earthquakes (blue dots). (**B**) The predicted epicenters from the FCN model trained with 1,013 historical events (red dots). (**C**) The epicenter error distribution of the testing earthquakes (red circles). (**D**) The depth error distribution of the testing earthquakes (red circles). The triangles denote the locations of the stations.

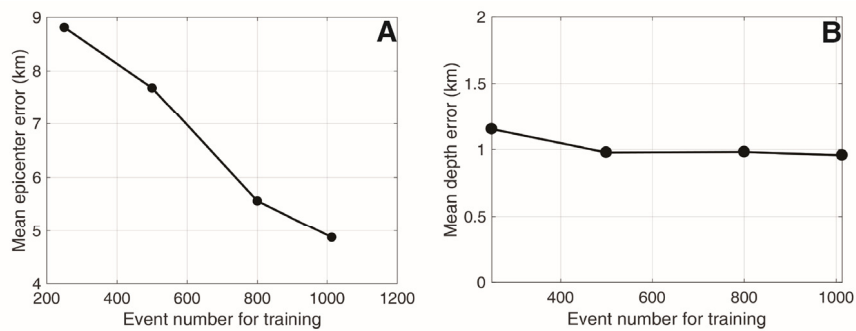

**Fig. 4. The effects of the size of the training set.** (**A**) The mean errors of the predicted epicenters by the FCN models trained with different numbers of training samples. (**B**) The mean errors of the predicted depth by the FCN models trained with different numbers of training samples.

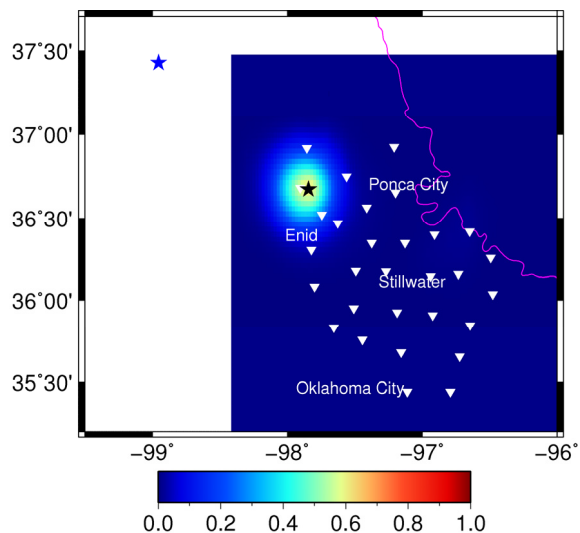

**Fig. 5. The predicted result obtained by inputting the waveform of an earthquake occurring outside the study area.** The maximum value of the Gaussian distribution for the earthquake is approximately 0.5, significantly less than those for the predicted events within the study area.

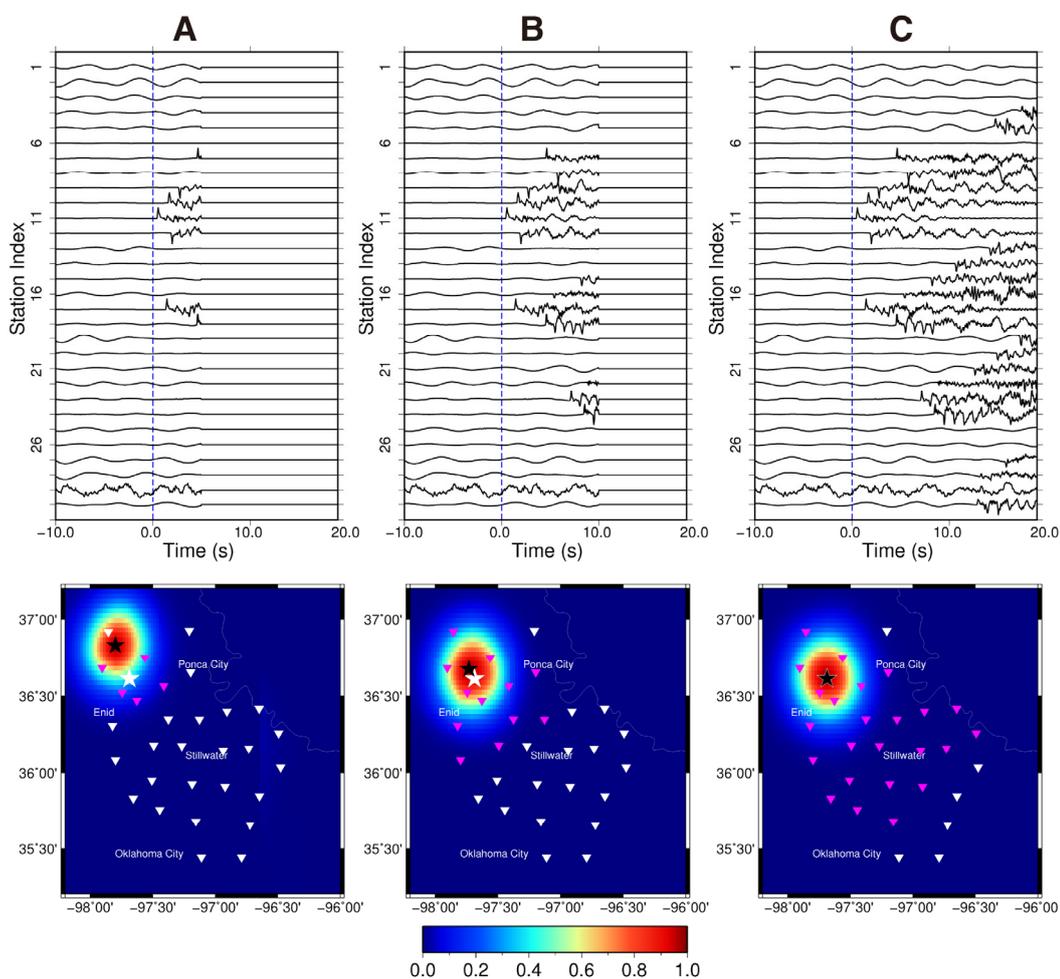

**Fig. 6. The potential application for earthquake early warning.** The location results predicted with input time windows of 5 s (**A**), 10 s (**B**), and 20 s (**C**). The purple triangles represent the stations receiving data, and the white triangles denote stations without data yet from the event; the black and white stars are the predicted and true locations, respectively; the color image represents the values of the Gaussian distribution.

## 3. DISCUSSION

We propose a fully convolution network to predict the hypocentral location of an earthquake in the form of a 3D probabilistic distribution image and apply the approach to monitor induced seismicity in Oklahoma. From testing 194 events with the FCN, the mean value of the epicenter location error is approximately 4.9 kilometers, and the depth error is approximately 1.0 km. Considering the station interval in the seismic network in this study varies from 14 to 30 km, we believe these results are fairly accurate. In addition, the FCN does not require a large amount of training set, it seems about 1,000 events from 30 stations are sufficient. Therefore, the approach could be applicable to areas with moderate to high seismicity or active seismic areas with a short period of instrumentation. The testing results show that an earthquake can be located reasonably well with partial initial data recorded at the first few stations, and the result can be further improved with the availability of additional data. This suggests great potential for the application of the proposed approach for earthquake early warning systems. In addition, our method utilizes network data, and testing shows that malfunction of individual stations may not affect the results. This is one of the advantages of utilizing data from a network over a single station.

The limitation of the method is the accuracy of ground truth and the coverage of the training set. The ground truth is derived from the conventional analysis of earthquake arrival times using a one-dimensional (1D) velocity model, therefore, the results may include errors due to oversimplified physics and low data quality. Improvement on the ground truth could be made by further applying relative location methods such as the double difference approach to minimize the influence of velocity heterogeneities (*12*). The number of training samples is important, but the coverage of the training samples in the interest zone is also essential. We observed that the large location errors are mainly in the areas without many training samples. The emerging compressive sensing technology could help reconstruct a dense data coverage from randomly located events (*38*). Combining synthetics to build a database may be another option (*23*).

In this study, we intend to solve the earthquake location problem by applying the fully convolutional network. This effort presents a new direction to solve geophysical inverse problems. The approach could be further applied to solve other geophysical problems, including velocity model building, statics solutions, elastic property inversion, and well-log interpretation.

## 4. METHODS

Our method constitutes a fully convolutional network (FCN) that takes in a window of three-component waveform data from multiple stations as volumetric input and predicts the earthquake location with a 3D image as the output. We propose the use of a 3D Gaussian distribution in the subsurface to delineate the probability distribution of an earthquake location, where each pixel represents a label with a probabilistic value. The peak position in the output volume represents the most likely earthquake location, and the magnitude of the peak value represents the probability of the result.

For each training event, we label the input data with the following Gaussian distribution:

$$\begin{cases} f(x,y,z) = \exp\left\{-\left[(x-x_0)^2 + (y-y_0)^2 + (z-z_0)^2\right]/r\right\} \\ x \in R_x, y \in R_y, z \in R_z \end{cases} \quad (1)$$

where $(x_0, y_0, z_0)$ denotes the location parameters of the earthquake, the ground truth is obtained from the U.S. Geological Survey earthquake catalog, $R_x$, $R_y$ and $R_z$ are the dimensional limits of the 3D zone of interest, and $r$ is the radius of the Gaussian function. To employ the FCN to locate a new event that is not included in the training dataset, the FCN is able to predict a 3D image, which may not exhibit a Gaussian distribution. However, the pixel with the largest peak value marks the event location.

### 4.1. Network architecture

Our network is mainly composed of convolutional layers, and the fully connected layer is abandoned in comparison with the convolutional neural network (CNN) classification method. The fully connected layer is commonly used as the final layer to output a vector of classification probabilities in image classification problems (*36*). However, for an earthquake location problem, thousands of classes are required if each location pixel is set as a class, and such a large-scale classification problem may require an enormous number of training samples to achieve an acceptable accuracy; unfortunately, the number of historical earthquakes is limited. Therefore, similar to the methods used in image segmentation (*28*), we choose to directly utilize the final convolutional layer to output a 3D volume of pixels representing the probability of an earthquake location.

The convolutional layer in the network architecture (Fig. 1A) is formulated as follows:

$$y_{ijc}^{l} = \sigma\left(\sum_{a=1}^{m}\sum_{b=1}^{n}\sum_{c'=1}^{C^l} y_{(i+a)(i+b)c'}^{l-1} \cdot w_{abc'}^{c}\right) \quad (2)$$

where $y^l$ is the output of the layer $l$; $w$ contains the weights for the filters in the current convolution layer; the output and input channels are indexed with $c$ and $c'$, respectively; the number of channels in layer $l$ is $C^l$; the kernel size of the filter is $m \times n$; and $\sigma$ is the nonlinear activation function. The input and output of the convolutional layer in equation 2 are both 3D arrays with dimensions of width, length and channel, and each channel of the layer output was obtained by convolving the channels of the previous layer with a bank of 2D filters applied in the width and length directions, as shown in equation 2. We utilize the zero-padded convolutional layer in the whole network; therefore, the width and length of the input are the same as those of the output. We utilize 21 convolutional layers in the neural network, as shown in Fig. 1A. The input of the network contains three-component waveform data, and each component corresponds to a channel of a color image simulating one of the RGB colors. The total number of seismic stations is 30, and the number of time samples of an event extracted from a data trace is 2048. Therefore, the input data are represented by a 3D volume (184,320 points) with dimensions of 2048 (time samples) by 30 (stations) by 3 (components). We set the kernel size $m \times n$ of all convolutional layers to be 3 × 3. The number of channels of features is increased from 3 to 1024 and then decreased from 1024 to 30. The output feature size of a convolutional layer is also determined by the number of channels in the current layer, and the weights $W$ are also related to the number of channels of the input feature according to equation 2. We utilize the rectified linear unit (ReLU) activation function in each layer, but the sigmoid function is utilized in the final convolution layer to output the final Gaussian distribution image.

In our application, we may also apply downsampling (maxpooling) and upsampling to the output of some of the intermediate convolution layers. The maxpooling layer is utilized to extract useful information from the waveform data. However, the input size (30×2048×3) is very different from the output size (80×128×30) in our application. The maxpooling and upsampling operations are also able to adjust the width and length of the features in the intermediate layers of the network, as shown in Fig. 1A. For example, we set the pooling size to be (1, 4) in the first maxpooling operation for the output of the 3rd convolutional layer, the width remains unchanged, and the length is decreased 4 times. The width and length of the features become 5×8 after the four maxpooling operations with pooling sizes of (1, 4), (1, 4), (2, 4), and (3, 4). To obtain the final location image, we utilize upsampling to increase the size of the features; the sizes for the three upsampling operations are (4, 4), (2, 2), and (2, 2), which means we repeat the rows and columns of the data by the two values of the size, respectively. Finally, the width and length of the features are increased to 80×128 after the three upsampling operations.

We output the 2D features by selecting eight channels from the 3D outputs in some of the layers to show how the waveform is transformed into the location image through each layer (Fig. 1A). Because the maxpooling layer is used to extract the features that are sensitive to the earthquake location, the features from the layers before the final maxpooling layer are similar to the input. However, the features are gradually transformed into the final Gaussian image as the layers become deeper.

## 4.2. Objective function of the training

We utilize a set of three-component waveform data from 30 seismic stations labeled with 3D probabilistic location images to train the network, and we adopt the binary cross-entropy loss function as follows:

$$\Psi = \frac{1}{N} \sum_{k=1}^{N} \sum_{d \in D} p_d^k \log(q_d^k) + (1 - p_d^k) \log(1 - q_d^k) \quad (3)$$

Where $p$ and $q$ are the predicted and true location image labels in this study; $N$ is the number of training samples; and $D$ is the assemblage of grid nodes in the location image. Because both the waveform data and the location image label require a substantial amount of memory for training, we minimize the loss function $\Psi$ using a batched stochastic gradient descent algorithm. The samples are shuffled prior to training, and then we divide the samples into several batches. At each training step, we feed the neural network a batch of samples and minimize the loss function to obtain the updated FCN model. This process is repeated until all samples are fed into the neural network, and then the current epoch is finished. In the application, we perform 200 epochs for all tests, and approximately 3 hours are required for training with 1013 events. After training the neural network, we apply the trained model to new data, which the model has not seen, and obtain an output consisting of a 3D probabilistic location volume.

We perform the tests based on TensorFlow (*39*), and the Adam algorithm is utilized to optimize the loss function for each batch at each epoch (*40*). The learning rate is set to 10-4, and the other parameters are set to the default values recommended by the authors of the Adam algorithm. We also utilize two dropout layers in the middle of the neural network architecture to regularize the training to avoid overfitting the data.

## 5. REFERENCES

1. L. Geiger, Probability method for the determination of earthquake epicenters from the arrival time only. *Bull. St. Louis Univ* **8**, 56-71 (1912).
2. W. Bakun, C. Wentworth, Estimating earthquake location and magnitude from seismic intensity data. *Bull. Seismol. Soc. Am.* **87**, 1502-1521 (1997).
3. P. G. Richards, F. Waldhauser, D. Schaff, W. Y. Kim, The Applicability of Modern Methods of Earthquake Location. *Pure Appl. Geophys.* **163**, 351-372 (2006).
4. W. L. Ellsworth, Injection-induced earthquakes. *Science* **341**, 1225942 (2013).
5. K. M. Keranen, M. Weingarten, G. A. Abers, B. A. Bekins, S. Ge, Sharp increase in central Oklahoma seismicity since 2008 induced by massive wastewater injection. *Science* **345**, 448-451 (2014).
6. F. R. Walsh, M. D. Zoback, Oklahoma's recent earthquakes and saltwater disposal. *Sci. Adv.* **1**, e1500195 (2015).

7. M. Weingarten, S. Ge, J. W. Godt, B. A. Bekins, J. L. Rubinstein, High-rate injection is associated with the increase in US mid-continent seismicity. *Science* **348**, 1336-1340 (2015).
8. C. Langenbruch, M. D. Zoback, How will induced seismicity in Oklahoma respond to decreased saltwater injection rates? *Sci. Adv.* **2**, e1601542 (2016).
9. B. R. Lienert, E. Berg, L. N. Frazer, HYPOCENTER: An earthquake location method using centered, scaled, and adaptively damped least squares. *Bull. Seismol. Soc. Am.* **76**, 771-783 (1986).
10. G. L. Pavlis, F. Vernon, D. Harvey, D. Quinlan, The generalized earthquake-location (GENLOC) package: an earthquake-location library. *Comput. Geosci.* **30**, 1079-1091 (2004).
11. W. Rodi, Grid-search event location with non-Gaussian error models. *Phys. Earth Planet. Inter.* **158**, 55-66 (2006).
12. F. Waldhauser, W. L. Ellsworth, A Double-Difference Earthquake Location Algorithm: Method and Application to the Northern Hayward Fault, California. *Bull. Seismol. Soc. Am.* **90**, 1353-1368 (2000).
13. S. D. Billings, Simulated annealing for earthquake location. *Geophys. J. Int.* **118**, 680-692 (1994).
14. B. Růžek, M. Kvasnička, Differential Evolution Algorithm in the Earthquake Hypocenter Location. *Pure Appl. Geophys.* **158**, 667-693 (2001).
15. H. Guo, H. Zhang, Development of double-pair double difference earthquake location algorithm for improving earthquake locations. *Geophys. J. Int.* **208**, 333-348 (2017).
16. G. Lin, The Source - Specific Station Term and Waveform Cross - Correlation Earthquake Location Package and Its Applications to California and New Zealand. *Seismol. Res. Lett.* **89**, 1877-1885 (2018).
17. S. J. Gibbons, F. Ringdal, The detection of low magnitude seismic events using array-based waveform correlation. *Geophys. J. Int.* **165**, 149-166 (2006).
18. C. E. Yoon, O. O'Reilly, K. J. Bergen, G. C. Beroza, Earthquake detection through computationally efficient similarity search. *Sci. Adv.* **1**, e1501057 (2015).
19. Z. Li, M. A. Meier, E. Hauksson, Z. Zhan, J. Andrews, Machine Learning Seismic Wave Discrimination: Application to Earthquake Early Warning. *Geophys. Res. Lett.* **45**, 4773-4779 (2018).
20. Z. E. Ross, M. A. Meier, E. Hauksson, T. H. Heaton, Generalized Seismic Phase Detection with Deep Learning. *Bull. Seismol. Soc. Am.* https://doi.org/10.1785/0120180080 (2018).
21. T. Perol, M. Gharbi, M. Denolle, Convolutional neural network for earthquake detection and location. *Sci. Adv.* **4**, e1700578 (2018).
22. A. C. Aguiar, G. C. Beroza, PageRank for earthquakes. *Seismol. Res. Lett.* **85**, 344-350 (2014).
23. J. Zhang, H. Zhang, E. Chen, Y. Zheng, W. Kuang, X. Zhang, Real-time earthquake monitoring using a search engine method. *Nat. Commun.* **5**, 5664 (2014).
24. U.S. Geological Survey, National Earthquake Information Center earthquake catalog; http://earthquake.usgs.gov/earthquakes/search/.
25. M. T. McCann, K. H. Jin, M. Unser, A review of convolutional neural networks for inverse problems in imaging. https://arxiv.org/abs/1710.04011 (2017).
26. L.-C. Chen, G. Papandreou, I. Kokkinos, K. Murphy, A. L. Yuille, Deeplab: Semantic image segmentation with deep convolutional nets, atrous convolution, and fully connected CRFs. *IEEE transactions on pattern analysis and machine intelligence* **40**, 834-848 (2018).
27. E. Shelhamer, J. Long, T. Darrell, Fully Convolutional Networks for Semantic Segmentation. *IEEE Transactions on Pattern Analysis and Machine Intelligence* **39**, 640-651 (2017).
28. O. Ronneberger, P. Fischer, T. Brox, U-net: Convolutional networks for biomedical image segmentation. https://arxiv.org/abs/1505.04597 (2017).
29. H. Zhang, T. Xu, H. Li, S. Zhang, X. Wang, X. Huang, D. Metaxas, StackGAN: Text to Photo-Realistic Image Synthesis with Stacked Generative Adversarial Networks, *in Proceedings of 2017 IEEE International Conference on Computer Vision (ICCV)*. (2017), Venice, Italy, 22 to 29 October 2017.
30. R. C. Alt, M. D. Zoback, In situ stress and active faulting in Oklahoma. *Bull. Seismol. Soc. Am.* **107**, 216-228 (2016).
31. M. D. Petersen, C. S. Mueller, M. P. Moschetti, S. M. Hoover, A. L. Llenos, W. L. Ellsworth, A. J. Michael, J. L. Rubinstein, A. F. McGarr, K. S. Rukstales, Seismic-hazard forecast for 2016 including induced and natural earthquakes in the central and eastern United States. *Seismol. Res. Lett.* **87**, 1327-1341 (2016).
32. K. M. Keranen, H. M. Savage, G. A. Abers, E. S. Cochran, Potentially induced earthquakes in Oklahoma, USA: Links between wastewater injection and the 2011 Mw 5.7 earthquake sequence. *Geology* **41**, 699-702 (2013).
33. C. Frohlich, W. Ellsworth, W. A. Brown, M. Brunt, J. Luetgert, T. MacDonald, S. Walter, The 17 May 2012 M4.8 earthquake near Timpson, East Texas: An event possibly triggered by fluid injection. *J. Geophys. Res.: Solid Earth* **119**, 581-593 (2014).
34. D. H. Sheen, A robust maximum - likelihood earthquake location method for early warning. *Bull. Seismol. Soc. Am.* **105**, 1301-1313 (2015).
35. H. C. Hsu, D. Y. Chen, T. L. Tseng, Y. M. Wu, T. L. Lin, H. C. Pu, Improving Location of Offshore Earthquakes in Earthquake Early Warning System. *Seismol. Res. Lett.* **89**, 1101-1107 (2018).
36. C. Szegedy, W. Liu, Y. Jia, P. Sermanet, S. Reed, D. Anguelov, D. Erhan, V. Vanhoucke, A. Rabinovich, Going deeper with convolutions, *in Proceedings of the IEEE conference on computer vision and pattern recognition.* (2015), Boston, MA, USA, 7 to 12 June 2015.

37. R. M. Allen, P. Gasparini, O. Kamigaichi, M. Bose, The status of earthquake early warning around the world: An introductory overview. *Seismol. Res. Lett.* **80**, 682-693 (2009).
38. E. Candès, J. Romberg, T. Tao, Robust uncertainty principles: Exact signal reconstruction from highly incomplete frequency information. *IEEE Trans. Inform. Theory* **52**, 489–509 (2006).
39. M. Abadi, A. Agarwal, P. Barham, E. Brevdo, Z. Chen, C. Citro, G. S. Corrado, A. Davis, J. Dean, M. Devin, S. Ghemawat, I. Goodfellow, A. Harp, G. Irving, M. Isard, Y. Jia, R. Jozefowicz, L. Kaiser, M. Kudlur, J. Levenberg, D. Mane, R. Monga, S. Moore, D. Murray, C. Olah, M. Schuster, J. Shlens, B. Steiner, I. Sutskever, K. Talwar, P. Tucker, V. Vanhoucke, V. Vasudevan, F. Viegas, O. Vinyals, P. Warden, M. Wattenberg, M. Wicke, Y. Yu, X. Zheng, TensorFlow: Large-scale machine learning on heterogeneous systems. https://arxiv.org/abs/1603.04467 (2015).
40. D. P. Kingma, J. Ba, Adam: A method for stochastic optimization. https://arxiv.org/abs/1412.6980 (2014).

**Acknowledgments: Founding:** We thank the financial support of National Natural Science Foundation of China (Grant No. 41674120 and 41704040), China Postdoctoral Science Foundation (Grant No. 2017M622010). **Author contributions:** X. Z. implemented the location software and conducted training and testing. J. Z. designed and supervised the project. C. Y., S. L., Z. C. and W. L. helped develop the method and analyze the results. J. Z. and X. Z. wrote the manuscript. All authors contributed ideas to the project. **Competing interests**: J. Z. is Board Director of Real Time Geomechanics and GeoTomo, geophysical technology companies. All other authors have no competing interests. **Data and materials availability**: The data used in this study can be requested from IRIS website: http://ds.iris.edu/ds/. The information of the seismic network is described by the website: http://www.fdsn.org/networks/detail/NX. The ground truth of the earthquake events are obtained from USGS website: http://earthquake.usgs.gov/earthquakes/search/

## Supplementary Materials

### Section S1. Selecting size of time window for the training set

To simulate applications for earthquake early warning, we define a training set with partial waveforms in different window lengths in an effort to predict the event location using partial data. The full event window for training is 90 s in length starting from the initial arrival of the event at the network. The partial event windows within 90 s could be set same for all training samples at the fixed lengths or varied for each event with random lengths selected. With 1,013 training events, we define new training samples by selecting three more time windows (fixed or random) for each event and setting the trace values after the length to zero. The fixed windows include data within 0-5 s, 0-30 s, 0-55 s and 0-90 s, and the total length of the traces is unchanged. The random windows are then selected between 0-90 s for each event. This effort augments training samples to 4,052 from 1,013. With fixed and random window selection, we created two FCN models, and both are trained with 4,052 samples.

To test the two FCN models, we select partial waveforms from 194 testing samples by keeping data within 8 time windows 0-5 s, 0-10 s, 0-20 s, 0-30 s, 0-45 s, 0-55 s, 0-70 s, and 0-90 s. That leads to a total of 1,552 tests with each FCN model. After testing, the epicenter errors are calculated for all of the results (fig. S1). Our primary concern is the epicenter errors. The comparison suggests that the epicenter errors with the random window size for the training set are consistently smaller than those with fixed window size. Four dash lines mark the fixed window lengths for the training set, suggesting accuracy lower even if the training set and testing sample are in the same data length. Random length for data selection seems preserving more information from data, which could be explained in terms of the compressive sensing and stochastic optimization (*38, 40*).

### Section S2: The location results with filtered data

Our FCN model takes raw data for the training set and testing sample. In a different effort, we want to study the performance of the network by using processed data. Specifically, we want to apply a bandpass filter to both the training and testing samples and observe the results. All of the 1,013 training samples and 194 testing samples are in 0-90s length. We apply a bandpass filter to the raw data and preserve data information between 2- 8 Hz. A comparison of raw data and filtered data of an event is shown in fig. S2.

Figure S3 shows the location results with filtered waveforms. The predicted location distribution is close to the ground truth, and the average errors are about 4.0 km in epicenter and 1.0 km in depth. The epicenter results are improved from 4.9 km of errors associated with raw data. We also compared the statistics of the errors between the testing results with filtered data and raw data (fig. S4). The comparison clearly shows the improvement for epicenter, but the depth errors remain the same, which are already small.

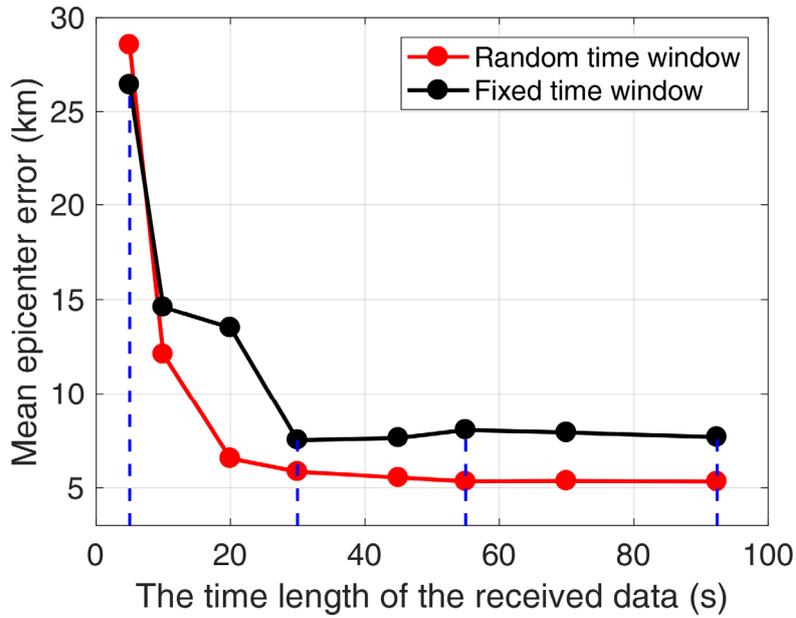

fig. S1. **The epicenter error comparison with random and fixed time window for the training set.** Two FCN models are created by the training set selected with fixed window lengths and randomly variable window lengths, respectively. The mean epicenter errors are calculated with 1,552 testing results for each FCN model. The dash lines mark the window length set for the fixed window tests.

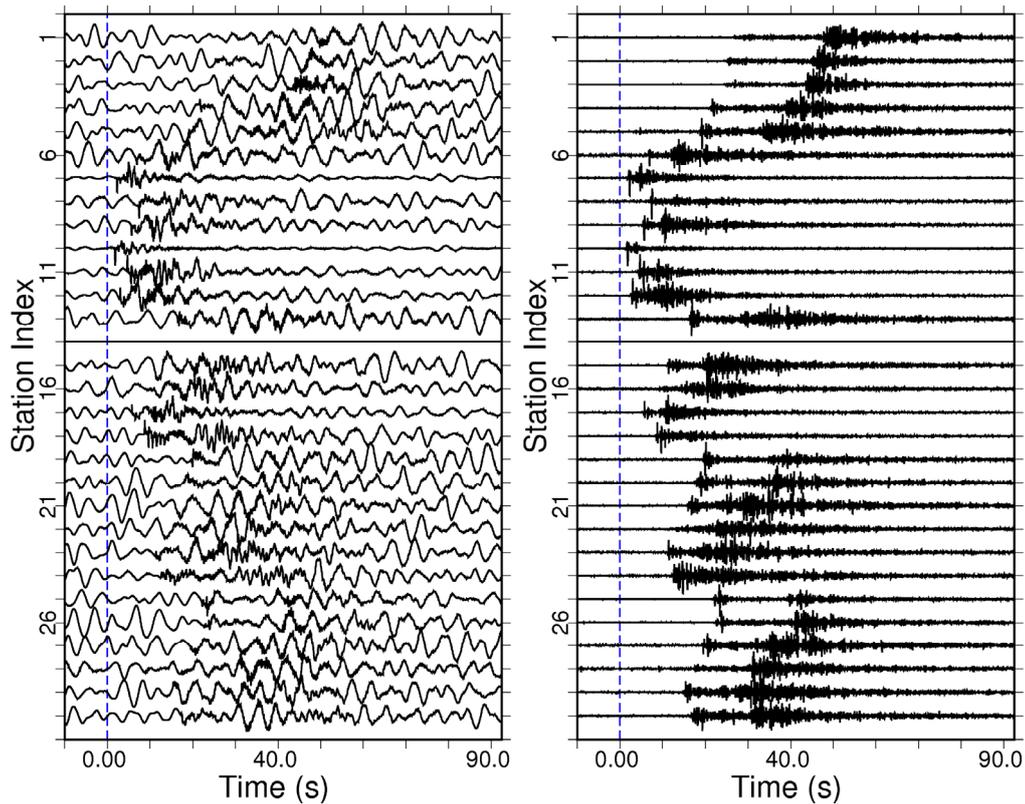

fig. S2. **The vertical component waveforms with (right) and without (left) filtering.** The right waveforms are bandpass filtered from 2 HZ to 8HZ.

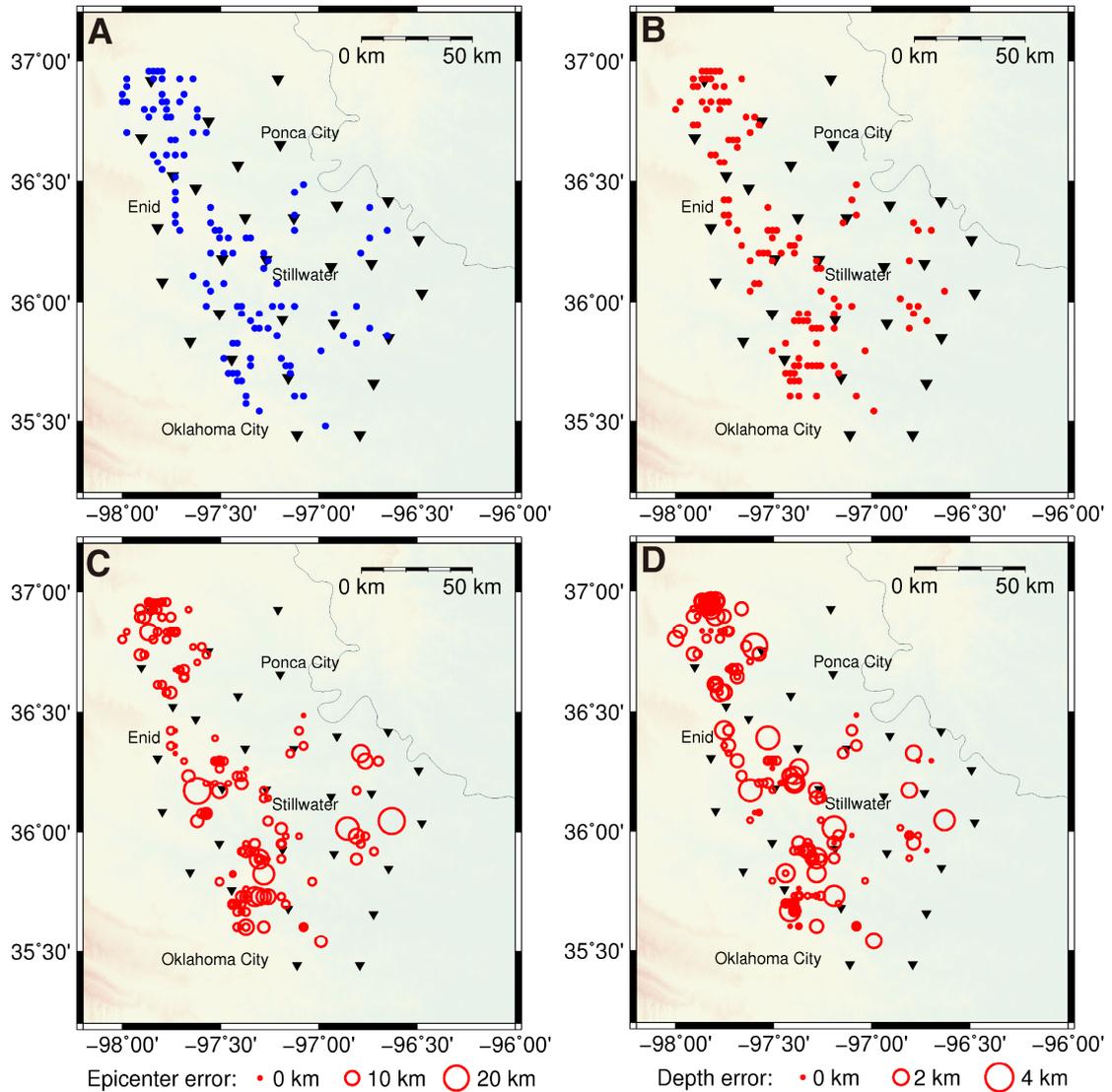

fig. S3. **The prediction results for 194 testing events with a bandpass filter applied.** (A) The ground truth of the 194 testing earthquakes; (B) the predicted epicenters from the FCN model trained with filtered events; (C) the epicenter error distribution of 194 testing events; (D) the depth error distribution of 194 testing events. The triangles denote the location of the seismic stations.

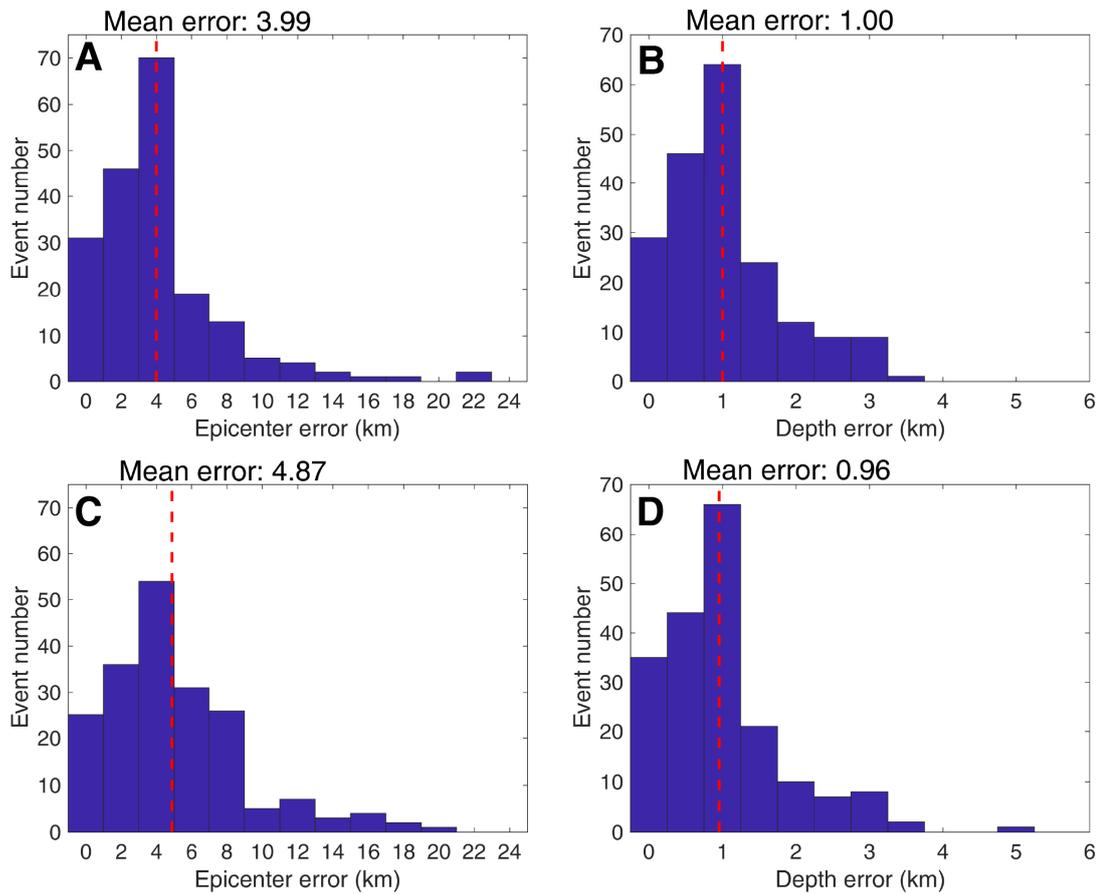

fig. S4. **The comparison of the error statistics for the testing with raw data or filtered data.** (A) The epicenter error statistics for 194 testing samples with filtered waveforms. (B) The depth error statistics for 194 testing samples with filtered waveforms. (C) The epicenter error statistics for 194 testing samples with raw waveforms. (D) The depth error statistics for 194 testing samples with raw waveforms. The red dash line marks the mean error of the total testing samples.